\documentclass[12pt]{article}
\usepackage[utf8]{inputenc}
\usepackage[T2A]{fontenc}
\usepackage[english]{babel}
\usepackage{color}
\usepackage{multicol}
\usepackage{amsmath, amssymb}
\usepackage{graphicx}
\usepackage{hyperref}
\usepackage[onehalfspacing]{setspace}
\usepackage[margin=2.5cm]{geometry}
\usepackage{upgreek}
\usepackage{amsmath}
\usepackage{lipsum}

\newcommand{\p}{\text{p}}
\newcommand{\q}{\text{q}}
\newcommand{\pmns}{U_{\rm PMNS}}
\newcommand{\Ol}[1]{\mathcal{O}\left(#1\right)}
\newcommand{\rank}[1]{{\rm rank}\left(#1\right)}
\newcommand{\diag}[1]{{\rm diag}\left(#1\right)}
\newcommand{\iss}[2]{{\rm ISS}\left(\text{#1},\text{#2}\right)}
\newcommand{\hc}{\text{H.c.}}
\newcommand{\re}[1]{{\rm Re}\left(#1\right)}
\newcommand{\im}[1]{{\rm Im}\left(#1\right)}

\begin{document}

\title{\bf Active-sterile neutrino mixing in the ISS(p,q) inverse seesaw models}
\author{M.N. Dubinin$^1$, E.Yu. Fedotova$^1$, D.M. Kazarkin$^{1,2}$}

\maketitle

\begin{center}
    \noindent $^1$ \textit{Skobeltsyn Institute of Nuclear Physics, Lomonosov Moscow State University, Moscow, Russia} \\
$^2$ \textit{Physics Department, Lomonosov Moscow State University, Moscow, Russia}

\begin{abstract}
\noindent 
A class of models with inverse seesaw (ISS) mechanism is considered for arbitrary numbers $\p$ and $\q$ of new neutral fermions for each generation of leptons. The models containing a candidate particle for the role of warm dark matter are sorted using the inversion technique of an arbitrary block matrix in terms of Schur complement. 
Studies of the neutrino mass hierarchies and active-sterile neutrino mixing scenarios in the ISS parameter space for the general case of Casas-Ibarra diagonalization make it possible to identify a class of models that satisfy the last strong constraints from NuSTAR, XMM-Newton, and eROSITA experiments for keV-scale sterile neutrino dark matter particles.
\end{abstract}

\end{center}

\section{Introduction}
\label{intro}

It is well-known that small masses of standard (active) neutrinos can be naturally generated by the seesaw type I mechanism \cite{mohapatra_0}, which uses three sterile right-handed Majorana neutrinos mixing with active left-handed Majorana neutrinos through the Yukawa interaction term. The masses of active neutrinos are determined by the relation $m_\nu \simeq - m_D \, M^{-1} m^T_D$, where $m_D$ is the 3$\times$3 mass matrix of the Dirac Lagrangian term and $M$ is the 3$\times$3 mass matrix of the Majorana lagrangian term with matrix elements much larger than the electroweak scale. Seesaw mechanism naturally arises in Grand Unified Theory (GUT) models. The simplest realizations are based on the ${\rm SO}(10)$ model \cite{Mohapatra:2006gs}, where fermion multiplets include the right-handed neutrinos along with the standard model (SM) fermions. For seesaw type I GUT typical values of $M_{ab}$ are of the order of $10^{10}$ GeV and higher, and the mixing effects are of the order of standard-to-heavy neutrino mass ratio, which makes the collider signals of the seesaw unobservable.   

A completely different framework is the inverse seesaw (ISS) mechanism \cite{mohapatra_2}, where two sets of singlet neutral fermions $\nu_R$ and $S_R$ are introduced for each generation of leptons $l_L$. In the framework of ${\rm SO}(10)$ GUT, the first of these sets is identified with the matter \textbf{16}-representation, while the other is a set of three singlet fermions. In the gauge basis $(\nu_L,\nu_R,S_R)$ after spontaneous symmetry breaking the mass term of the neutrino is given by   

\begin{equation}
     \frac{1}{2}\left(
        \begin{array}{ccc}
            \overline{\nu_L} & \overline{\nu_R^c} & \overline{S_R^c}
        \end{array} 
    \right)
    \left(
        \begin{array}{ccc}
            0 & m_D & 0 \\ 
            m^T_D & 0 & M_R \\
            0 & M^T_R & \mu 
        \end{array} 
    \right) 
    \left(
        \begin{array}{c}
            \nu_L^c \\
            \nu_R \\
            S_R
        \end{array}
    \right) + \hc,
\end{equation}
which corresponds to the Lagrangian
\begin{equation}
\label{lagr}
    -\mathcal{L}^{\rm ISS}_{\rm mass} = Y({\bar l}_L {\tilde \Phi}) \nu_R + M_R \overline{\nu^c_R} S_R + \frac{\mu}{2} {\bar S}^c_R S_R + \hc ,
\end{equation}
where $l_L$ is a left SM lepton doublet, $\Phi$ is a SM Higgs doublet with vacuum expectation value $\langle \Phi \rangle = \frac{v}{\sqrt{2}}$, $\tilde{\Phi} = \epsilon_{ij} \Phi^\dagger$ ($ \epsilon_{ij}$ -- antisymmetric tensor), 
$Y$, $M_R$, $\mu$ are complex matrices and $m_D=Y\langle \Phi \rangle$. Moreover, matrix $\mu$ is symmetric ($\mu^T = \mu$). If the lepton number breaking scale $||\mu||$ is sufficiently small\footnote{Here we do not specify the choice of the matrix norm $||\cdot||$, as it is not essential for our analysis. For example, one can use the Frobenius norm $||A||_F \equiv \sqrt{{\rm tr}(AA^\dagger)}$.} in comparison with $||M_R||$, then the left-handed neutrinos can acquire very small masses at the sub-eV scale, consistent with experimental data on active neutrino oscillations \cite{pdg2024}. Within the framework of nonsupersymmetric GUT, other realizations of the inverse seesaw mechanism were considered \cite{gut_iss_1, gut_iss_2}. In contrast to  the seesaw type I, the ISS mechanism naturally provides singlet neutrinos at the GeV or TeV scale with enhanced effects of lepton flavor violation \cite{Abada:2012cq}.

Seesaw mechanism involving an intermediate-mass state appeared 
under the name ''incomplete seesaw'' \cite{Johnson:1986ux, Glashow:1990dg, Gelmini:1994az}. A detailed analysis of inverse seesaw models with varying field content \cite{Abada:2012cq, Abada:2014kba, Abada_2017} and, notably, classification of scenarios featuring a light sterile neutrino as a dark matter candidate was carried out in \cite{Lucente_2014, Abada_wdm}.
In addition to the presence of candidate particles for the role of dark matter, various extensions by sterile neutrinos make it possible to generate baryon asymmetry of the Universe by means of thermal leptogenesis \cite{Fukugita:1986hr}. Lepton number violating decays of heavy neutral leptons (HNL's) produce the lepton asymmetries which are then converted into baryon asymmetries by means of sphaleron interactions \cite{Kuzmin:1985mm}.
Suppression of the SM neutrino masses by the $\mu$ parameter of the mass matrix in the ISS framework, where the masses of HNL's are low and almost degenerate, gives a natural realization of the leptogenesis scenario. These circumstances draw attention to general $\iss{p}{q}$ scenarios \cite{Abada_2017}, the consequences of which may differ significantly de\-pen\-ding on the numbers $(\p,\q)$ of neutral fermions with Majorana mass terms involved.

 The current experimental constraints 
 from searches for keV-scale dark matter
 are not satisfied under the typical naive seesaw estimate of the mixing $U^2_{\rm DM} \propto \frac{||m_\nu||}{||\mu||}$ which turns out to be too large. Further an improved approach to evaluate the mixing parameters using a generalization of the Casas-Ibarra parameterization for the $\iss{p}{q>p}$ model with a mass matrix of the form \eqref{eq:full-M-pq} is being developed. This allows one to build a consistent $\iss{2}{3}$ model that simultaneously reproduces the expected results from oscillation experiments and satisfies X-ray constraints by introducing the special hierarchy 
 in the Yukawa interaction of $\nu_R$ and $S_R$. 
The repository with the Mathematica code for our numerical calculations is available at \url{https://github.com/KazarkinD/ISS-pq-mixing}.

The paper is organized as follows. In Section \ref{sec:diagonalization} we consider the general case of analytical diagonalization of neutrino mass matrix with non-square blocks using the Schur complement. The Casas-Ibarra parametrization is discussed there to ensure the observed mass differences for active neutrinos. 
The obtained results are investigated analytically in Section \ref{sec:toy-models} within toy models $\iss{1}{1}$ and $\iss{1}{2}$. 
The simplest seesaw model with viable warm DM,  $\iss{2}{3}$, is analyzed numerically in Section \ref{sec:iss23}.
Results and discussion are summarized in Section \ref{sec:results}. 
The appendix contains a brief description of the used elements of linear algebra.

\section{Analytical diagonalization and active-sterile mixing}
\label{sec:diagonalization}

\subsection{Inverse seesaw mass matrix 
\label{subsec:iss-mass-matrix}
}

A key assumption for constructing the model is the possibility of extension of the SM neutrino sector by two types of
neutral fermions. The complete set of neutral fermion fields is given by $(\nu_{L,{\alpha}}, \nu_{R,a}, S_{R,b})$, where $\alpha=e,\mu,\tau $ and indices for nonstandard particles are $a=\overline{1,\p}$ and $b=\overline{1,\q}$.

In the following we will employ the notation  
\begin{equation}  
    \label{eq:full-M-pq}
    \mathbf{M}_{(\p,\q)} = 
        \begin{pmatrix}
            \mathbb{O}_{3\times3} & m_{D~3\times \p} & \mathbb{O}_{3 \times \q}  \\
            m^T_{D~\p\times3} & \mathbb{O}_{\p\times \p} & M_{R~\p\times \q} \\
            \mathbb{O}_{\q\times3} & M^T_{R~\q\times \p} & \mu_{~\q\times \q}    
        \end{pmatrix},
        \quad 
        \Psi \equiv 
        \begin{pmatrix}
            \nu^c_{L\,\alpha} \\
            \nu_{R\,a} \\
            S_{R\,b}
        \end{pmatrix}.
\end{equation}
Here, $\mathbb{O}$ denotes a matrix with all elements equal to zero. A key object in this model is the matrix $\mu$ corresponding to a small Majorana mass term that breaks the full lepton number by $\Delta L = 2$. 
Then the Lagrangian \eqref{lagr} can be rewritten in a concise form   
\begin{equation}
\label{lagr-iss}
     -\mathcal{L}^{\rm ISS}_{\rm mass} = \frac{1}{2}
     \overline{\Psi^c} \, \mathbf{M}_{(\p,\q)} \, \Psi + \hc 
\end{equation}
This model will be referred to as $\iss{p}{q}$ in the following. Unitary transformation of states $\Psi = \mathbf{U} \Psi^\prime$ is used to diagonalize the mass matrix.

One can rewrite the mass matrix in the seesaw type I-like form \cite{Ibarra:2010xw}
\begin{equation}
    \mathbf{M}_{(\p,\q)} = 
    \begin{pmatrix}
        \mathbb{O}_{~3 \times 3} & \widetilde{m}_{D~3\times (\p+\q)} \\
        \widetilde{m}^T_{D~(\p+\q)\times 3} & \mathcal{X}_{~(\p+\q)\times (\p+\q)}
    \end{pmatrix},
\end{equation}
where the designation $\cal{X}$ for the sterile neutrino matrix block 
\begin{equation}
    \mathcal{X} = \begin{pmatrix}
        \mathbb{O}_{~\p \times \p} & M_{R~\p\times \q} \\
        M^T_{R~\q\times \p} & \mu_{~\q \times \q}
    \end{pmatrix}
\end{equation}
is used and $\widetilde{m}_D\equiv (m_D, ~\mathbb{O})$. The naturalness condition $||\mu|| \ll ||m_D|| \ll ||M_R||$ allows one to assume that $\mathcal{X}$ is unchanged after the first step of diagonalization\footnote{The same situation occurs for the diagonalization of canonical seesaw type I, when $M_M \simeq M_N + \Ol{\theta m_D}$, where $M_N = U_N^T \diag{M_1,M_2,M_3} U_N$. See e.g. \cite{own_jetp, own_sym} for details about seesaw type I diagonalization.} (see details in Section \ref{subsec:sterile-mass}).

The complete mass matrix $\mathbf{M}_{(\p,\q)}  \equiv \mathbf{M}$ of the size $(3+\p+\q)\times(3+\p+\q)$ is symmetric and, in general, complex valued. So, the Autonne-Takagi (AT) decomposition \cite{Horn_Johnson_book} is valid, $\mathbf{U}^T \mathbf{M} \mathbf{U} = \diag{m_1, \dots, M_1, \dots}$,
where we choose $\mathbf{U}$ transformation as a composition of two transformations $\mathbf{U} = \mathbf{W} \cdot \mathbf{V}$ \cite{Ibarra:2010xw},
\begin{eqnarray}
\label{ci}
    \mathbf{W} &=& \exp{(\omega)}\simeq \mathbf{I} +\omega + \dots,
    \\
    \mathbf{V} &=& 
        \begin{pmatrix}
            U_{\nu~3\times3}^* & \mathbb{O}_{3\times (\p+\q)} \\
            \mathbb{O}_{(\p+\q)\times3} & ~~\mathcal{U}_{~(\p+\q)\times(\p+\q)}
        \end{pmatrix}, 
\end{eqnarray}
corresponding to two steps of diagonalization. Here the antihermitian matrix $\omega$ ($\omega = - \omega^\dagger$) due to unitarity of $\mathbf{U}$ can be parametrized as 
\begin{equation*}
    \omega = 
        \begin{pmatrix}
            \mathbb{O}_{3\times 3} & -\tilde{\theta}_{3\times(\p+\q)} \\
            \tilde{\theta}^\dagger_{(\p+\q)\times 3} & ~~\mathbb{O}_{(\p+\q)\times (\p+\q)}
        \end{pmatrix},
        \quad 
    \tilde{\theta} = \left( \underset{3\times \p}{\theta_{1}} \, ,\, \underset{3\times \q}{\theta_{2}} \right).
\end{equation*}

Note that the exponent of the anti-Hermitian matrix \eqref{ci} in the framework of seesaw diagonalization was introduced in Ref.~\cite{Ibarra:2010xw} 
where it was assumed that $||\tilde{\theta}|| < I$. This approximation is very useful for obtaining analytical representations of mass matrices and mixings (see Section \ref{sec:diagonalization}). However, the numerical estimates carried out below (see Section~\ref{subsec:num-analysis} and the supplemental materials) also take into account
the explicit exponential representation for the matrix $\mathbf{W}$, Eq.~\eqref{ci}, the results are in good agreement.

Analytical diagonalization procedure is performed in the first order approximation for the case of a small mixing and is carried out in two stages: at the first stage the effective mass operator for active neutrinos $m_\nu \equiv U_\nu \hat{m} U_\nu^T$  is obtained, where for brevity we have introduced the notations $\hat{m} \equiv \diag{m_1, m_2, m_3}$. 
Furthermore, the chosen parameterization \eqref{ci} gives $U_\nu \simeq \pmns$ with $\Ol{\tilde\theta\tilde\theta^\dagger}$ accuracy ($\pmns$ is the Pontecorvo-Maki-Nakagava-Sakata mixing matrix). 
At the second stage the sterile sector matrix $\mathcal{X}$ is diagonalized.

\subsection{Active neutrino mass matrix}
\label{subsec:act-mass}

Similar to the case of the type I seesaw mechanism, in the case of inverse seesaw the masses are acquired by 'integrating out' the heavy fields at a scale below the electroweak scale.
The block-diagonalization procedure for the active neutrino mass operator $m_\nu$ yields the following form
\begin{equation}
    \label{eq:seesaw-iii-a}
    m_\nu = - \widetilde{m}_D \mathcal{X}^{-1} \widetilde{m}_D^T,
\end{equation}
which, as can be easily seen, is the Schur complement (see \cite{shur_compl_1, shur_compl_2} and also Eq. \eqref{def:shur-compl} in the Appendix) for the matrix $\mathbf{M}$ usually denominated as $\left( \mathbf{M} | \mathcal{X}\right)$ . In order to obtain the inverse matrix $\mathcal{X}^{-1}$, we use the block matrix inversion formula (the Frobenius formula, see Eq. \eqref{f:froben-inverse} in the Appendix)
\begin{equation}
\label{eq:inverse-sterile-block}
    \mathcal{X}^{-1} = 
    \begin{pmatrix}
        (\mathcal{X}|\mu)^{-1} & -(\mathcal{X}|\mu)^{-1} M_R \mu^{-1} \\
        -\mu^{-1} M_R^T  (\mathcal{X}|\mu)^{-1} & ~\mu^{-1} + \mu^{-1}M_R^T (\mathcal{X}|\mu)^{-1} M_R \mu^{-1}
    \end{pmatrix}.
\end{equation}
Then, using Eqs.~\eqref{eq:seesaw-iii-a} and \eqref{eq:inverse-sterile-block} we get the inverse seesaw formula of the known structure for $\iss{p}{q}$\footnote{ A similar expression for the case of a square matrix $m_D$ was used in the literature, see e.g. \cite{Abada_2017}.  }
\begin{equation}
\label{eq:seesaw-iii}
   m_\nu = U_\nu \hat{m} U_\nu^T 
    = m_D \left(M_R \mu^{-1} M_R^T\right)^{-1} m_D^T.
\end{equation}

In accordance with the experimental data on the difference of the squares of the neutrino masses $\Delta m^2_{ij}$ \cite{pdg2024},
the resulting form of the diagonal mass matrix for active neutrinos must include at least two distinct non-zero masses. In order for Eq.\eqref{eq:seesaw-iii} to be respected, it is necessary that the ranks of the left-hand and right-hand sides of this equation coincide, which imposes a constraint on the values of $\p$ and $\q$
\begin{equation}
\label{eq:p-q-limit}
    \min{(\p,\q,3)}     
    = 2\,\text{ or }\,3.
\end{equation}
When $\p > \q$, the formal representation of expression \eqref{eq:seesaw-iii} becomes incomputable because the
matrix 
$(\mathcal{X}|\mu) = - M_R \mu^{-1} M_R^T$
becomes invertible. Indeed, if $\p > \q$, then for a $\p \times \p$ matrix $(\mathcal{X}|\mu)$ we have $\rank{\mathcal{X}|\mu} = \q$. It follows in turn that the matrix $(\mathcal{X}|\mu)$ is not invertible (or, in other words, singular). On the other hand, numerical analysis shows that it is impossible to reproduce the masses and mixing parameters of the active neutrinos when $\p > \q \geq 2$.
So in the following we will consider the case of $\p \leq \q$. The compliance with experimental data can be achieved if $\q \geq \p \geq 2$ in the $\iss{p}{q}$, and for $\p = 2$ the model describes only two massive active neutrinos, 
while the third one remains massless within the inverse seesaw mechanism. For $\p \geq 3$, the model describes three distinct non-zero masses of active neutrinos. 
This scenario is naturally realized when the emergence of a third non-zero eigenvalue for the active neutrino mass matrix is related to the presence of a non-zero Majorana mass term of the form $\frac{1}{2} \overline{\nu_L^c}m_L\nu_L + \hc$, for instance, due to a non-standard Higgs sector \cite{Duka:1999uc} or loop contributions to neutrino masses \cite{Grimus:1989pu}. In that case, the relation \eqref{eq:p-q-limit} should be interpreted as the specific tree-level contribution of the $\iss{p}{q}$ mechanism to the active neutrino masses. However, within the scope of this work, we will not delve into the details of the case $m_L \neq 0$.

\subsection{Casas-Ibarra parametrization}

\label{subsec:casas-ibarra}
Diagonalization of the neutrino mass matrix $m_\nu$ performed using the matrix $U_\nu \simeq \pmns$ should lead to a result consistent with 
the data on $\Delta m_{ij}^2 \equiv m_{i}^2 - m_{j}^2$ \cite{pdg2024}.
Although the available cosmological constraints on the sum of the neutrino masses $\sum m_i$ are model-dependent, they indicate that the mass of the lightest neutrino is very small $m_{\rm light} \ll \Delta m_{ij}^2$ \cite{planck-2018}.
Thus a meaningful set for neutrino masses can be chosen as
\begin{equation}
    \begin{array}{cccc}
         m_1 = m_{\rm light}, & m_2 = \sqrt{|\Delta m_{21}^2|}, & m_3 = \sqrt{|\Delta m_{31}^2|} & \text{ for NH}; \\
         m_1 = \sqrt{|\Delta m_{31}^2|}, & m_2 = \sqrt{|\Delta m_{32}^2|}, & m_3 = m_{\rm light} & \text{for IH},
    \end{array}    
\end{equation}
where normal and inverted mass hierarchies for active neutrinos are denoted as NH and IH, correspondingly.  
Note that choosing $m_{\rm light} = 0$ allows for a self-consistent use of the $\iss{2}{3}$ model to describe neutrino masses. This model will be discussed in detail in Section \ref{sec:iss23}.
Construction of an explicit form of the Yukawa matrix $Y_D \equiv \sqrt{2}\frac{m_D}{v}$ using known experimentally measured parameters $m_i$ and $\pmns$ was proposed in \cite{Ibarra:2010xw} and is known in the literature as Casas-Ibarra (CI) parametrization. In the context of inverse seesaw models, the CI parameterization 
has been constructed for the mass matrix in the form of \eqref{eq:seesaw-iii-a} with all square blocks (a number of papers has considered the case $\p=\q=3$). In the case under consideration, CI parameterization is constructed for a lower-dimensional matrices (Schur complement $\left(\mathcal{X}|\mu\right)$) with $\p \neq \q$ using the form of Eq. \eqref{eq:seesaw-iii}. For brevity, let's denote $S \equiv-\left(\mathcal{X}|\mu\right)$. 
AT decomposition of a symmetric square matrix $S$ has the form $S = U_S \hat{S} U_S^T$, where $\hat{S}$ is the diagonal matrix with singular values of $S$. Using Eq. \eqref{eq:seesaw-iii} with $\p < \q$, one can obtain 
\begin{eqnarray}
\label{eq:ci-details-1}
    U_\nu \sqrt{\hat{m}}\sqrt{\hat{m}} U_\nu^T &=& m_D U_S^* \sqrt{\hat{S}^{-1}} \sqrt{\hat{S}^{-1}} U_S^\dagger m_D^T ,
    \\
\label{eq:ci-details-2}
    \Omega &=& \sqrt{\hat{m}^{-1}} U_\nu^\dagger m_D U_S^* \sqrt{\hat{S}^{-1}},
    \\
\label{eq:ci-details-3}
    m_D &=& U_\nu \sqrt{\hat{m}} \Omega \sqrt{\hat{S}} U_S^T.
\end{eqnarray}
For different values of $\p$, the $\Omega$ matrix has different forms
\begin{eqnarray}
\label{omega:p=2}
    \p = 2: &\qquad & \Omega = 
    \begin{pmatrix} 
    0~~0 \\
    \mathcal{R}_{2\times 2} 
    \end{pmatrix}_{\rm NH}
    ~ 
    \text{ or } \quad
    ~
    \begin{pmatrix} 
        \mathcal{R}_{2\times 2} \\
        0~~0    
    \end{pmatrix}_{\rm IH},
    \quad 
    \mathcal{R}_{2\times 2} \in O(2, \mathbb{C}) ;
    \\
\label{omega:p=3}
    \p = 3: &\qquad & \Omega = \mathcal{R}_{3\times 3}, \quad \mathcal{R}_{3\times3} \in O(3, \mathbb{C});
    \\
\label{omega:p>3}
    \p > 3: &\qquad & \Omega = \mathcal{R}_{3 \times \p}, \quad \mathcal{R}\mathcal{R}^T = I_{3\times 3}.
\end{eqnarray}
Thus, a possible scheme for CI parametrization in the $\iss{p}{q>p}$ model can be implemented as follows 
\begin{enumerate}
    \item[1)] the AT decomposition $S=U_S \hat{S} U_S^T$ for $\p \times \p$ matrix $S \equiv M_R \mu^{-1}M_R^T$ is formed;
    \item[2)] the mass hierarchy and active neutrino mass ordering $m_i$ is chosen, $\hat{m}=\diag{m_1,m_2,m_3}$. When $\p = 2$, the lightest neutrino's mass $m_{\rm light}$ should be set to zero;
    \item[3)] the Dirac mass matrix $m_D$ is calculated using  
    \begin{equation}
        \label{eq:ci-md}
        m_D^{\rm (CI)} \simeq \pmns \sqrt{\hat{m}} \Omega \sqrt{\hat{S}} U_S^T;
    \end{equation}
    \item[4)] $\Omega$ matrix is selected according to conditions \eqref{omega:p=2}-\eqref{omega:p>3}. 
\end{enumerate}
The full ISS Majorana mass matrix \eqref{eq:full-M-pq} constructed using this scheme is guaranteed to contain the prescribed masses of active neutrinos among its nonzero singular values, and the mixing of active neutrinos is described by the PMNS matrix. 

\subsection{Active-sterile mixing parameters}
\label{subsec:mixing}

The small mixing between active and sterile states can be encoded into $\tilde{\theta}$ matrix, where $\tilde{\theta} \simeq \widetilde{m}_D\mathcal{X}^{-1}$, or, separately for each mixing blocks,
\begin{eqnarray}
\label{eq:theta-1}
    \theta_1 &\simeq & - m_D \left(M_R \mu^{-1} M_R^T \right)^{-1}, \\
\label{eq:theta-2}
    \theta_2 &\simeq & -\theta_1 M_R \mu^{-1}.
\end{eqnarray}

Within the CI parametrization, these matrices take the forms
\begin{eqnarray}
\label{eq:theta-1-ci}
    \theta_1 
     & \simeq & -U_{\rm PMNS} \sqrt{\hat{m}} \Omega \sqrt{\hat{S}^{-1}} U_S^\dagger
    \\
\label{eq:theta-2-ci}
    \theta_2 &\simeq &  U_{\rm PMNS} \sqrt{\hat{m}} \Omega \sqrt{\hat{S}^{-1}} U_S^\dagger M_R \mu^{-1}.
\end{eqnarray}

In general, mixing $\theta_2$ is enhanced compared to $\theta_1$ by a significant factor $M_R \mu^{-1}$.
Naive estimate based on the dimensional analysis with model scales gives 
$S \propto \frac{||M_R||^2}{||\mu||}$, 
so
\begin{equation}
\label{eq:ci-naive-estimate}
    m_D \propto ||M_R|| \sqrt{\frac{m_{\rm act}}{||\mu||}} ,
\qquad 
\theta_1 \propto \frac{\sqrt{m_{\rm act} \cdot||\mu||}}{||M_R||} ,
\qquad 
\theta_2 \propto \sqrt{\frac{m_{\rm act}}{||\mu||}} ,
\end{equation}
where we use the estimation of the active neutrino masses as $0.01 \text{ eV} < m_{\rm act} < 0.1 \text{ eV}$. It is interesting to point out that with CI parametrization
the $\theta_2$ component is independent of $M_R$-scale changes. This observation is verified numerically in the $\iss{2}{3}$ model, see Sections \ref{sec:iss23} and \ref{sec:results}. 

\subsection{Mass matrix of the sterile sector}
\label{subsec:sterile-mass}

The second step involves diagonalization transformations performed separately for the active and sterile blocks (using matrices $U_\nu$ and $\mathcal{U}$, respectively), namely, $U_\nu ^\dagger m_\nu U_\nu^* = \hat{m}$ and $\mathcal{U}^T \mathcal{X} \mathcal{U} = \hat{\mathcal{X}}$, where $\hat{\mathcal{X}} \equiv \diag{M_1, \dots, M_{\p+\q}}$.

It is worth noticing that the theoretical estimates were carried out under the assumption $\mathcal{X} \simeq \mathcal{X}_{\rm w}$, where $\mathcal{X}_{\rm w}$ is a $(\p+\q)\times(\p+\q)$ matrix in the flavor basis obtained by converting the matrix $\mathbf{M}_{(\p,\q)}$ to a block-diagonal form
\begin{equation}
    \mathbf{W}^T \mathbf{M}_{(\p,\q)} \mathbf{W}  = \left(
    \begin{array}{cc}
    m_\nu & 0 \\
    0 & \mathcal{X}_{\rm w} 
    \end{array}
    \right).
\end{equation}
In the second order of  decomposition by the $\tilde{\theta}$, the matrix of heavy neutral leptons can be represented as
\begin{equation}
    \mathcal{X}_{\rm w} \simeq \mathcal{X} + \frac{1}{2} \left( 
    \tilde{\theta}^\dagger \tilde{\theta} \mathcal{X} + \mathcal{X}^T \tilde{\theta}^T \tilde{\theta}^* \right).
\end{equation}
Using the explicit forms of the mixing matrices \eqref{eq:theta-1},  \eqref{eq:theta-2}, one can get
\begin{equation}
    \mathcal{X}_{\rm w} \sim
    \begin{pmatrix}
        0 & M_R \\
        M_R^T & \mu
    \end{pmatrix}
    +
    \begin{pmatrix}
        -m_D^2 M_R^{-2} \mu & m_D^2 M_R^{-3} \mu^2 \\
        m_D^2 M_R^{-1} & m_D^2 M_R^{-2} \mu
    \end{pmatrix},
\end{equation}
which shows that within the considered scale hierarchy (see Subsection~\ref{subsec:iss-mass-matrix}), the approximation $\mathcal{X} \simeq \mathcal{X}_{\rm w}$ is valid.
 
The non-trivial task of the second step
is to find the form of the transformation $\mathcal{U}$. In the case $\p = \q$, the matrix $M_R$ is square and therefore a block rotation by $\frac{\pi}{4} \cdot \mathbf{1}_{p\times p}$ can be used. If, however, $\p < \q$, then the spectrum of eigenvalues contains $\q-\p$ small quantities corresponding to light Majorana states, as well as to $2\p$ heavy Majorana states, which form $\p$ \textit{pseudo-Dirac} states \cite{pseudo-dirac}. Let us illustrate the statement above with two characteristic toy-models $\iss{1}{1}$ and $\iss{1}{2}$. 

\section{Some toy-models}
\label{sec:toy-models}

\subsection{$\iss{1}{1}$}
\label{subsec:toy-iss11}

The simplest model considered is the “toy model” $\iss{1}{1}$. This model has the characteristic feature that, in addition to light active neutrinos, it only contains heavy \textit{pseudo-Dirac pair} in its particle spectrum. Sterile mass block of Majorana mass matrix $\mathbf{M}_{(1,1)}$ has the simple form
\begin{equation}
    \mathcal{X} =
        \begin{pmatrix}
            0 & M \\
            M & \mu
        \end{pmatrix}, \qquad \mu \ll M,
\end{equation}
the eigenvalues of the mass matrix are given by
\begin{equation}
    \lambda_{1,2} = \frac{1}{2}\left( \mu \pm \sqrt{\mu^2 + 4M^2} \right) \simeq \frac{\mu}{2} \pm M,
\end{equation}
and the diagonalizing transformation is
\begin{equation}
    \mathcal{U} =
    \begin{pmatrix}
        \frac{M}{\sqrt{\lambda_1^2 + M^2}} & \frac{M}{-\sqrt{\lambda_2^2 + M^2}} \\
        \frac{\lambda_1}{\sqrt{\lambda_1^2 + M^2}} & \frac{\lambda_2}{-\sqrt{\lambda_2^2 + M^2}}
    \end{pmatrix}
    \simeq \frac{1}{\sqrt{2}}
    \begin{pmatrix}
        1 & -1 \\
        1 & 1
    \end{pmatrix}.
\end{equation}
Correct sign of $N_2$ mass can be achieved by redefinition $N_{2R} \to +i N_{2R}$, which is equivalent of redefining the corresponding Majorana field $N_2 \to i \gamma_5 N_2$.
The connection between the gauge (or flavor) basis $\Psi$ and the mass basis $\Psi^\prime = (\upnu_L^c, \, N_{1R},\,N_{2R})^T$ (see Sec.\ref{sec:diagonalization}) provides important information about the interaction strength of new particles with the SM particles. In particular, the flavor states of active neutrinos $\nu_\alpha$ are a mixture of the form
\begin{equation}
    \nu_{L\,\alpha} \simeq P_L \big[ (U_\nu)_{\alpha i} \,\upnu_{i} - (\theta^*_1)_{\alpha} N_{+} - (\theta^*_2)_\alpha N_{-} \big],  
\end{equation}
where $N_{\pm} = \frac{1}{\sqrt{2}}\left( N_1 \pm iN_2\right)$ are pseudo-Dirac fields, since the Majorana fields $N_1$ and $N_2$ are almost degenerate in mass. It should be noted that $(N_{-})^c = N_{+}$. From Eqs.~\eqref{eq:theta-1} and \eqref{eq:theta-2} we get 
\begin{equation}
    (\theta_1)_\alpha = - (m_D)_\alpha \frac{\mu}{M^2}, \qquad (\theta_2)_\alpha = (m_D)_\alpha \frac{1}{M} .
\end{equation}
It follows that the naturalness condition $\mu \ll ||m_D|| \ll M$ leads to the fact that the flavor neutrino is a mixture of massive active neutrino states with a small admixture of the $N_{-}$ particle state. At the same time, the admixture of the $N_{+}$ antiparticle state is vanishingly small and suppressed by a factor of $\mu / M$ relative to the $N_{-}$ admixture.

\subsection{$\iss{1}{2}$}
\label{subsec:toy-iss12}

An illustrative example of an inverse seesaw model with a viable warm dark matter candidate can be a model with $\p = 1$ and $\q = 2$, see \cite{Abada_wdm}. However, as noted above, this model cannot generate more than one massive active neutrino and is therefore classified as a toy model. The primary matrix of the sterile sector is
\begin{equation}
    \mathcal{X} =
    \begin{pmatrix}
        0 & M_1 & M_2 \\
        M_1 & \mu_1 & 0 \\
        M_2 & 0 & \mu_2 \\
    \end{pmatrix},
\end{equation}
where $M_1,M_2$ are "large" components of the mass matrix ($M_1 \sim M_2$), and $\mu_1,\mu_2$ are ''small'' 
parameters ($\mu_i \ll M_j$).

Let us represent the matrix as the sum of the unperturbed part and the small perturbation $\mathcal{X} = \mathcal{X}_0 + \delta \mathcal{X}$, where  
\begin{equation}
    \mathcal{X}_0 = 
    \begin{pmatrix}
        0 & M_1 & M_2 \\
        M_1 & 0 & 0 \\
        M_2 & 0 & 0 \\
    \end{pmatrix}, 
    \quad 
    \delta \mathcal{X} = 
    \begin{pmatrix}
        0 & 0 & 0 \\
        0 & \mu_1 & 0 \\
        0 & 0 & \mu_2 \\
    \end{pmatrix}.
\end{equation}
The eigenvalues $\lambda_i$ corresponding to eigenvectors $\mathbf{v}_i$ are searched in the form
\begin{equation}
\lambda \simeq \lambda^{(0)} + (\mathbf{v}_i^{(0)})^\dagger \delta \mathcal{X} \mathbf{v}_i^{(0)}, \quad \text{where } \quad 
\mathcal{X}_0 \mathbf{v}_i^{(0)} = \lambda_i^{(0)} \mathbf{v}_i^{(0)}.
\end{equation} 
Here we use unperturbed vectors and the perturbation matrix to find corrections 
in the first order of perturbation theory, see also \cite{Abada_wdm}.

First-order eigenvalues and orthonormal zero-order eigenvectors are 
\begin{eqnarray}
    \lambda_1 & \approx & \frac{M_2^2 \mu_1 + M_1^2 \mu_2}{M_0^2}, 
    \quad
    \mathbf{v}_1^{(0)} = \frac{1}{M_0} 
    \begin{pmatrix} 
        0 \\ 
        M_2 \\ 
        -M_1 
    \end{pmatrix},
    \nonumber 
    \\
    \lambda_{2,3} & \approx & \pm M_0 + \frac{M_1^2 \mu_1 + M_2^2 \mu_2}{2M_0^2},
    \quad 
    \mathbf{v}_{2,3}^{(0)} = \frac{1}{\sqrt{2} M_0} 
    \begin{pmatrix}
        M_0 \\
        \pm M_1 \\ 
        \pm M_2
    \end{pmatrix},
\end{eqnarray}
where $M_0 \equiv \sqrt{M_1^2 + M_2^2}$.
Then the transformation matrix of sterile states to the mass basis with zero-order accuracy has the form 
\begin{equation}
    \mathcal{U}^{(0)} = 
    \frac{1}{M_0}
    \begin{pmatrix}
        0 & \frac{M_0}{\sqrt{2}} & \frac{M_0}{\sqrt{2}} \\
        -M_2 & \frac{M_1}{\sqrt{2}} & -\frac{M_1}{\sqrt{2}} \\
        M_1 & \frac{M_2}{\sqrt{2}} & - \frac{M_2}{\sqrt{2}}
    \end{pmatrix}.
    \label{eq:U0}
\end{equation}
Now, let us isolate the asymptotic dependence on the inverse seesaw scale. If we assume that $\mu_1 = \mu_2 \equiv \mu_{DM}\sim\Ol{\text{keV}}$, and $M_1 = M_2 \equiv M \gg \mu_{\rm DM}$, then
\begin{eqnarray}
    m_{N_1} &=& |\lambda_1| \simeq \mu_{DM} \sim {\cal O}(\text{keV}), \nonumber \\
    m_{N_2} &=& |\lambda_2| \simeq \sqrt{2}M + \frac{\mu_{DM}}{2} , \nonumber \\
    m_{N^\prime_3} &=& |\lambda_3| \simeq \sqrt{2}M - \frac{\mu_{DM}}{2}   , \\
    \mathcal{U} &=& 
    \begin{pmatrix}
        0 & \frac{1}{\sqrt{2}} & \frac{1}{\sqrt{2}} \\
        - \frac{1}{\sqrt{2}} & \frac{1}{2} & -\frac{1}{2} \\
        \frac{1}{\sqrt{2}} & \frac{1}{2} & -\frac{1}{2}
    \end{pmatrix}. \label{eq:iss-1-1-mixing} 
\end{eqnarray}
The matrix \eqref{eq:iss-1-1-mixing} is known as Bi-Maximal mixing \cite{Jarlskog:1998uf}. Here we use the redefinition of $N_3 \to i\gamma_5 N_3$ as in the previous model to get the correct sign of the mass term. The flavor states of active neutrinos can be presented as 
\begin{eqnarray}
    \nu_{L~\alpha} &\simeq& P_L \Bigg( 
    (U_\nu)_{\alpha i} \upnu_i 
    + \Big[ 
        \frac{(\theta_2^*)_{\alpha2} - (\theta_2^*)_{\alpha1}}{\sqrt{2}}
        \Big] 
        N_1 
    - 
    \Big[
        \frac{(\theta_{2}^*)_{\alpha1} + (\theta_{2}^*)_{\alpha2}}{\sqrt{2}} 
        \Big] 
        N_{+}
       -
       (\theta_{1}^*)_{\alpha}  N_{-}  \Bigg), \quad
\end{eqnarray}
where $N_\pm = \frac{1}{\sqrt{2}} \left( N_2 \pm i N_3 \right)$.
The mixing parameters of the flavored left-handed neutrino $\nu_{L\alpha}$ with the light sterile particle $N_1$ are determined by the values of the matrix $|\theta_{2,\alpha i}|^2$. 
Admixture of $N_{-}$ is significantly suppressed in comparison with $N_{1}$ and $N_{+}$.

In the general case ($\mu_1 \neq \mu_2$ and $M_2 \neq M_1$), the mixing matrix of the lightest HNL with active neutrinos has the form
\begin{equation}
    \mathbf{U}_{\alpha 4} \simeq (m_D^*)_\alpha \frac{M_1 M_2}{M_0} \frac{\mu_2 - \mu_1}{M_1^2 \mu_2+M_2^2 \mu_1},
\end{equation}
where the unitary transformation of zero-order accuracy \eqref{eq:U0} is taken into account. 
In the special case $M_2 = \epsilon M_1$ ($\epsilon \ll 1$), one can note that the mixing matrix
\begin{equation}
    \mathbf{U}_{\alpha 4} \simeq (m_D^*)_\alpha \frac{\epsilon}{M_1} 
    \left(1 - \frac{\mu_1}{\mu_2}\right)
    \label{eq:ThetaISS12}
\end{equation}
depends linearly on the parameter $\epsilon$. 

\section{Warm dark matter in $\iss{2}{3}$}
\label{sec:iss23}

\subsection{Light sterile neutrino in $\iss{2}{3}$}
\label{subsec:iss23-light-N}

A minimal realistic model describing data on neutrino oscillations, containing two nonzero masses of active neutrinos, and containing one light candidate particle for the role of warm DM with a mass of 
${\cal O}$(keV) is the $\iss{2}{3}$ \cite{Lucente_2014, Abada_wdm}.
Such mass is achieved by postulating the scale of the Majorana mass parameter $||\mu||\sim\Ol{10\text{ keV}}$.
A significant simplification at this step is to use the diagonal form for the $\mu$ matrix later on,
$\mu = \hat{\mu} = {\rm diag}\left( \mu_1, \mu_2, \mu_3 \right)$. This assumption does not limit the generality (at least as long as we do not consider the gauge model with $S$ fields, see Eq. (\ref{lagr}), in non-singlet representations, therefore the choice of their initial basis is arbitrary). Then
\begin{equation}
    \label{eq:X-iss-2-3}
    \mathcal{X} = 
        \begin{pmatrix}
            0 & 0 & M_{11} & M_{12} & M_{13} \\
            0 & 0 & M_{21} & M_{22} & M_{23} \\
            M_{11} & M_{21} & \mu_1 & 0 & 0  \\
            M_{12} & M_{22} & 0 & \mu_2 & 0  \\
            M_{13} & M_{23} & 0 & 0 & \mu_3
        \end{pmatrix},
\end{equation}
where fulfillment of the naturalness condition $\mu_i \ll M_{kj}$ ($i,j = \overline{1,3}$ and $k=1,2$) is implied. In order to obtain the mass spectrum of heavy sterile states, let's consider the "unperturbed" mass matrix $\mathcal{X}_0$ which is obtained from $\mathcal{X}$ by switching off small Majorana mass terms, so $\mu_i=0$. It is easy to show that the natural assumption about full-ranked matrix $M_R$ ($\rank{M_R} =2$) implies that the $\rank{\mathcal{X}_0} =4$. This, in turn, means that $\mathcal{X}_0$ matrix contains four nonzero eigenvalues corresponding to the four heavy states.
The remaining one zero eigenvalue of the matrix $\mathcal{X}_0$ corresponds to a small eigenvalue of the matrix $\mathcal{X}$ which is of the order of $\Ol{\mu}$. Thus, the spectrum of $\iss{2}{3}$ contains $3-2=1$ light neutral lepton $N_1$ (which is a candidate for DM particle) with the mass $m_{N_1} \equiv m_{DM} \sim \Ol{\mu} $ and $(3+2) - 1 = 4$ heavy Majorana neutral leptons, grouped into two pseudo-Dirac pairs. 

\begin{figure}
    \centering
    \includegraphics[width=0.49\linewidth]{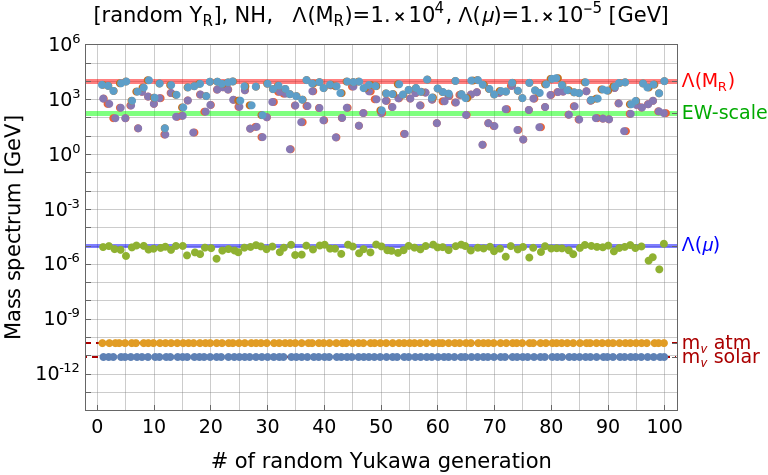}
    \includegraphics[width=0.49\linewidth]{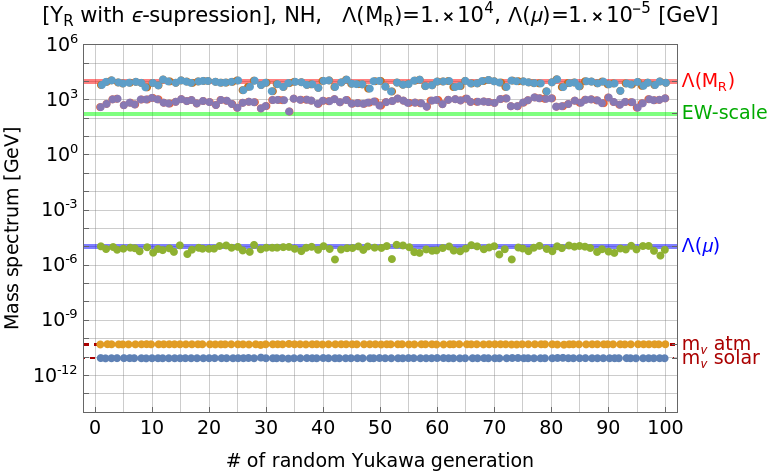}
\caption{ Numerical simulation of the mass spectrum for the $\iss{2}{3}$ model with a viable warm Dark Matter candidate at the keV scale for diagonal complex matrix $\mu = \Lambda(\mu) Y_\mu$ (uniform distribution, $z_{\min}=0.1+i\cdot0.1,$ $z_{\max} = 1 + i$, see Eqs. \eqref{eq:random-uniform},\eqref{eq:random-log-uniform}). 
    Here $m_{\nu,~{\rm atm}} = \sqrt{|\Delta m_{21}^2|}$, $m_{\nu~{\rm solar}} = \sqrt{|\Delta m_{32}^2|}$, where $\Delta m_{ij}^2 $ are differences of squared masses of the active neutrinos from oscillation data \cite{pdg2024}, $\Lambda(M_R)$ and $\Lambda(\mu)$ denote the scales of the matrix blocks, $M_R$ and $\mu$. CI parametrization \eqref{eq:ci-md} is used with $\mathcal{R}_{2\times2} = I_{2\times2}$, normal hierarchy. \textbf{Left panel:} random Yukawa matrix $Y_R$ with log-uniform generation, $z_{\min} = 10^{-6} + i \cdot 10^{-6}$ and $z_{\max} = 1 + i $. \textbf{Right panel:} Yukawa matrix $Y_R$ with 
    special dark matter Yukawa pattern \eqref{eq:YR-from-ansatz}.   
    }
    \label{fig:mass-spectrum}
\end{figure}

\subsection{Numerical analysis}
\label{subsec:num-analysis}
In this section we present the calculation results for the mixing matrix using a numerical AT decomposition method based on SVD matrix decomposition \cite{svd_method} which is applied to a symmetric mass matrix $\mathbf{M}_{(2,3)}$. The numerical SVD method is implemented in the built-in procedure \texttt{SingularValueDecomposition[]} of the \texttt{Wolfram Mathematica 14.0} package \cite{wolfram_SVD}.
In the following two denominations for the dimensionful parameters $\Lambda(M_R)$ and $\Lambda(\mu)$ are introduced which correspond to the scales of the matrix blocks $M_R \equiv \Lambda(M_R) Y_R$ and $\mu = \Lambda(\mu) Y_\mu$. For the numerical analysis complex Yukawa matrices $Y_X$, $X=\{R,\,\mu\}$ are generated using random complex numbers inside the rectangular\footnote{The notation $\left[z_{\min}, z_{\max}\right]$ is a rectangular region in the complex plane, where $z_{\min}$ is the coordinate of the lower-left corner and $z_{\max}$ is the coordinate of the upper-right one.} on the complex plane with two options
\begin{eqnarray}
\label{eq:random-uniform}
    \text{uniform generation:} &\quad& 
    \begin{array}{c}
         (Y_X)_{mn} = A_{mn} + iB_{mn}, \\
          A_{mn} \in [\re{z_{\min}}, \re{z_{\max}}],
          \\ 
           B_{mn} \in [\im{z_{\min}}, \im{z_{\max}}],
    \end{array} 
    \\
    \label{eq:random-log-uniform}
    \text{log-uniform generation:} &\quad& 
    \begin{array}{c}
        (Y_X)_{mn} = \exp{(a_{mn})} + i \exp{(b_{mn})}, \\
        a_{mn} \in [\re{\ln{z_{\min}}}, \re{\ln{z_{\max}}}], \\
         b_{mn} \in [\im{\ln{z_{\min}}}, \im{\ln{z_{\max}}}],
    \end{array}
\end{eqnarray}
where the components of matrices $A,B, a, b$ are (pseudo)random real numbers with uniform distribution in the specified intervals. 

In order to verify numerically the qualitative observations made in the previous Section \ref{subsec:iss23-light-N} we employ a numerical procedure to determine the singular value spectrum of the full mass matrix for the $\iss{2}{3}$ model. 
The spectra were obtained for two cases: the first case with a random matrix $Y_R$ with elements which are log-uniform
within the range $|(Y_R)_{mn}| \sim 10^{-6} - 1$ with the boundaries $z_{\min} = 10^{-6} + i\cdot 10^{-6}$ 
and $z_{\max} = 1 +i\cdot 1$, and the second case of special hierarchy for dark matter scenario, Eq.\eqref{eq:ansatz} and \eqref{eq:YR-from-ansatz}, which   will be discussed in detail below.
The results of 100 random trials of the Yukawa matrices with fixed dimensionful scales are shown in Fig.~\ref{fig:mass-spectrum}.
In both cases the masses of active neutrinos generated correspond to experimental data \cite{pdg2024}.
The Majorana mass scale of $S_R$ fields consistently accounts for the dark matter mass. 
The masses of the four remaining heavy leptons form two nearly degenerate pairs with a mass scale for each pair set by the $\nu_R - S_R$ Dirac mass term. 
As expected, numerical simulations exhibit a larger spread in predictions when the elements of $Y_R$ matrix are drawn using a log-uniform distribution (see Fig.~\ref{fig:mass-spectrum}, left panel) compared to a uniform distribution (Fig.~\ref{fig:mass-spectrum}, right panel)\footnote{This reduction in spread for the uniform case is a direct consequence of a smaller dynamical range of random numbers generation. }.
Note that the right figure corresponds to a specific dark matter-inspired pattern (see the text and Eqs.\eqref{eq:x-ray-problem}--\eqref{eq:YR-from-ansatz} below), while the left figure shows purely random case. 
The comparison reveals that the DM scenario yields characteristic splitting of the two pseudo-Dirac heavy states: one remains at the $M_R$ scale, while the other drops by about one order of magnitude for $\Lambda(M_R) = 10$ TeV. 
The overall form of these dependencies is similar for the inverted hierarchy of active neutrino masses, except for the spectrum of active neutrinos, which are generated, as expected, only near $m_{\nu~{\rm atm}}$.
Numerical results for the mixing parameters of the light sterile neutrino $N_1$ are used to obtain the mixing parameter 
\begin{equation}
    \label{eq:uudm-mixing}
    U_{\rm DM}^2 \equiv \sum_{\alpha=e,\mu,\tau} |\mathbf{U}_{\alpha 4}|^2,
\end{equation}
which can be reconstructed experimentally in searches for $N_1$ decays essential for DM phenomenology \cite{vysotsky, xray-Wolfenstein, xray-Abazajian}
\begin{equation}
    \label{eq:sterile-width}
    \Gamma_{N_1 \to 3\nu} = \frac{G_F^2 m_{\rm DM}^5}{96\pi^3} U^2_{\rm DM}, 
    \quad 
    \Gamma_{N_1\to \gamma \nu} = \frac{27 \alpha_{\rm EM}}{4\pi} \Gamma_{N_1\to 3\nu}.
\end{equation}
The choice of the fourth column in Eq.~(\ref{eq:uudm-mixing}) is based on neutrinos ordering within the SVD-decomposition by increasing eigenvalues of the mass matrix, so that
the lightest sterile neutrino $N_1$ is the next mass eigenstate after the three active ones. 
One of the major challenges of describing neutral heavy lepton dark matter is to ensure that the mixing $U^2_{\rm DM}$ has an appropriate size in accordance with the available data from gamma-ray astronomical telescopes NuSTAR \cite{nuSTAR}, XMM-Newton \cite{XMM} and eROSITA \cite{erosita} (and also data from future projects like Athena \cite{athena}). According to the available gamma-ray astronomical data, within the mass range of sterile neutrino dark matter 2 keV$< m_{\rm DM} <$10 keV it is meaningful to require that $U^2_{\rm DM} \lesssim 10^{-10}$ (more precise constraints depend on the specific value of $m_{\rm DM}$). In this case a naive approximate estimate of mixing of the order of magnitude $U^2_{\rm DM} \propto \left(\frac{v}{\Lambda(M_R)}\right)^2$ in combination with the masses of active neutrinos $||m_\nu|| \propto \left(\frac{v}{\Lambda(M_R)}\right)^2 \Lambda(\mu)$  gives a naive criteria
\begin{equation}
\label{eq:x-ray-problem}
    U^2_{\rm DM} \cdot \Lambda(\mu) \simeq ||m_\nu|| \simeq 0.01\text{ eV}-0.1 \text{ eV},  
\end{equation}
which obviously fails for the required values of $\Lambda(\mu) \simeq m_{\rm DM} \sim 1-10 \text{ keV}$ and $U^2_{\rm DM} \lesssim 10^{-10}$. 

To solve the problem of inconsistency, it is proposed to introduce a hierarchy of Yukawa constants $Y_R$. Based on the results of a numerical experiment\footnote{with fixed $\mathcal{R}_{2\times2} = I$ and diagonal $\mu$} of random generation of Yukawa matrices, which uses threshold-based filtering for $U^2_{\rm DM}$ value, a general template $F(\epsilon)$ for the internal hierarchy of the elements of the $Y_R$ matrix was constructed
\begin{equation}
\label{eq:ansatz}
    F(\epsilon) = 
        \begin{pmatrix}
            1 & 0.1 & \epsilon 
            \\
            0.1 & 0.01 & \epsilon
        \end{pmatrix}, 
\end{equation}
which contains a small parameter $\epsilon$ that ensures the desired suppression of the mixing term $U^2_{\rm DM}$ for $\epsilon \leq 10^{-4}$.

\begin{figure}[htbp]
\centering
  \begin{minipage}[t]{0.48\textwidth}
    \centering
    \vspace{0pt}
    \includegraphics[width=\linewidth]{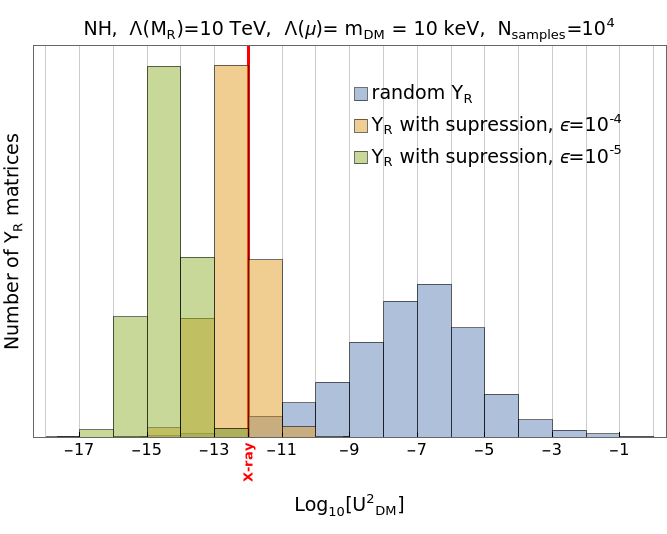} 
    \caption{ Histogram showing the distribution of the order of magnitude of $U^2_{\rm DM}$ mixing parameter \eqref{eq:uudm-mixing}, calculated for a sample of 10 000 randomly generated Yukawa matrices for three scenarios: (1) random $Y_R$ matrix (log-uniform, $z_{\min}=10^{-6} + i 10^{-6}$, $z_{\max}=1+i$); (2) $Y_R$ matrix with an element hierarchy template $F(\epsilon)$ with $\epsilon = 10^{-4}$, see Eqs.~\eqref{eq:ansatz}, \eqref{eq:YR-from-ansatz}; (3) The same as scenario (2) but with $\epsilon = 10^{-5}$. Normal hierarchy and $\mathcal{R}_{2\times2} = I$ for Casas-Ibarra parametrization are used in all three scenarios.}
    \label{fig:hist}
  \end{minipage}
  \hfill
  \begin{minipage}[t]{0.48\textwidth}
    \centering
    \vspace{0pt}
    \includegraphics[width=\linewidth]{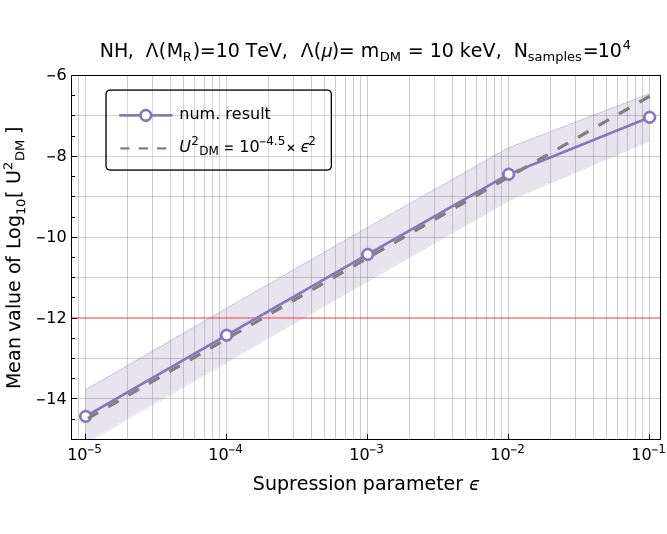} 
    \caption{
    The dependence of the mixing parameter on the suppression parameter $\epsilon$ from the hierarchy template for $Y_R$, obtained by averaging the parameter over 10 000 randomly generated Yukawa matrices with template \eqref{eq:ansatz} according to scheme \eqref{eq:YR-from-ansatz}. The thickness of the purple band corresponds to one standard deviation. The gray dashed line represents an approximate power-law dependence $U^2_{\rm DM} \sim \epsilon^2$. Horizontal redline denotes X-ray bound $U^2_{\rm DM} < 10^{-12}$ for $m_{\rm DM} = 10$ keV \cite{nuSTAR}.
    }
    \label{fig:logUUdm-Eps}
  \end{minipage}
\end{figure}

The Yukawa matrix $Y_R$ is formed as
\begin{equation}
\label{eq:YR-from-ansatz}
    (Y_R)_{mn} = \mathcal{C} F(\epsilon)_{mn} y_{mn}, 
\end{equation}
where $\mathcal{C}$ is a common numerical factor\footnote{
$\mathcal{C}$  can be less (or even much less) than 1. It does not affect on the mixing $\epsilon-$suppression, but reduces the mass spectrum of pseudo-Dirac HNLs.
} for the entire matrix, $y$ is an arbitrary dimensionless complex-valued matrix with $|y_{mn}| \sim 1 $. 
In our numerical simulation $y_{mn}$ are random complex elements with log-uniform distribution and $z_{\min}=0.1 + i\cdot0.1$, $z_{\max}=1 + i$.
Fig. \ref{fig:hist} shows a histogram for the distribution of the generated Yukawa matrices as a function of the logarithm of the dark matter mixing parameter, based on a set of 10 000 sample points. The property of suppressing $U^2_{\rm DM}$ mixing remains the same if a matrix with columns rearranged relative to the original $F$ is used as a template. Thus, there is a family of matrices $\{F\text{ with all column permutations}\}$ corresponding to the scenario with a viable sterile dark matter neutrino in the model $\iss{2}{3}$ ($\iss{2}{3}$-DM scenario).

Comparisons of precision are appropriate between the analytical block diagonalization with $\mathbf{W}$-transformation (see Section \ref{sec:diagonalization}) and a numerical method applied straightforwardly to diagonalize the entire mass matrix $\mathbf{M}_{(2,3)}$ in the form of Eq.\eqref{eq:full-M-pq}. 
For random Yukawa matrices using log-uniform generation with $z_{\min} = 10^{-6} + i \cdot 10^{-6}$, $z_{\max} = 1 + i$ and $\Lambda(M_R) = 10$ TeV, $\Lambda(\mu) = 10$ keV, the relative error $||\Delta||$ is of the order of $ 10^{-6}$.
Here $\Delta \equiv |\mathbf{U}_{\rm (num)}| - |\mathbf{U}_{\theta}|$, where $\mathbf{U}_{\rm (num)}$ is the mixing matrix evaluated using the fully numerical method; $\mathbf{U}_{\theta}$ is the mixing matrix calculated using symbolic diagonalization in the first step  and a subsequent numerical procedure for the evaluation of the matrix $\mathcal{U}$.


\begin{figure}[htbp]
    \centering
    \includegraphics[width=0.7\linewidth]{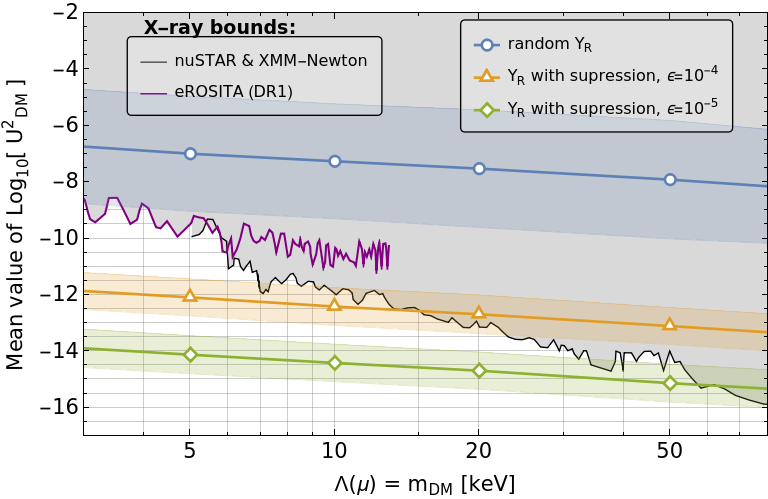}
    \caption{Values of the dark matter mixing parameter in the $\iss{2}{3}$ model, averaged over a random sample of 1000 Yukawa matrices, for different scales of the Majorana mass term $\Lambda(\mu)$. The width of the bands corresponds to one standard deviation of the value $\log{U^2_{\rm DM}}$. The three scenarios shown here are the same as those in Fig. \ref{fig:hist}. Purple and gray lines correspond to X-ray upper bounds from data of eROSITA \cite{erosita} (Data Release 1, 95\% CL) and nuSTAR \cite{nuSTAR} \& XMM-Newton \cite{XMM} combined data (95\% CL). To compare the results with existing constraints on the sterile dark matter neutrino, the condition $\Lambda(\mu) = m_{\rm DM}$ is used.}
    \label{fig:UUdm-Mdm}
\end{figure}

\section{Results and discussion}
\label{sec:results}

The article considers the general case of models $\iss{p}{q}$ with an inverse seesaw mechanism for generating neutrino masses.
The proposed diagonalization transformation of the ISS mass term in the neutrino sector, known in the literature for other seesaw mechanisms as Casas-Ibarra diagonalization \cite{Ibarra:2010xw}, explicitly involves the expected masses of the active neutrinos and the active neutrino mixing matrix $\pmns$.
Analytical results for construction of the mixing matrix are generalized
to the case $\p\neq \q$, matrix structures for $\theta_1$ and $\theta_2$
(Section \ref{subsec:mixing}) allow to analyze mixing hierarchies in a clear and unambiguous way. The proposed diagonalization requires operations with matrices of lower dimension, $\left(\mathcal{X}|\mu\right)$ instead of $\mathcal{X}$, due to the use of the Schur complement \eqref{def:shur-compl}, which allows for more efficient and accurate computations and, for some models, yields analytical solutions when required.

The article also contains a detailed analysis of diagonalization by numerical methods for the selected model $\iss{2}{3}$ with a viable dark matter candidate, as a result of which a characteristic form for the Yukawa matrices $Y_R$ is found, Eqs.~\eqref{eq:ansatz}, \eqref{eq:YR-from-ansatz}, allowing to suppress the phenomenological dark matter mixing parameter below the level of recent $X$-ray astronomical constraints by NuSTAR, XMM-Newton and eROSITA orbital observatories consistently with the available data on active neutrino masses and mixings. A required suppression of the mixing parameter holds even if the matrix $Y_R$ differs from the template $\eqref{eq:YR-from-ansatz}$ by any permutation of columns. Numerical analysis (see Fig.~\ref{fig:logUUdm-Eps}) demonstrates the relation $U^2_{\rm DM} \propto \epsilon^2$.
This situation is illustrated analytically within the toy-model $\iss{1}{2}$, see Eq.~\eqref{eq:ThetaISS12}.
The required order of magnitude is already achieved at $\epsilon = 10^{-4}$, and robust agreement over a wide mass range of $2 \text{ keV}<m_{\rm DM}<50 \text{ keV}$ is achieved at $\epsilon = 10^{-5}$, see Fig.~\ref{fig:UUdm-Mdm}.

An important observation, consistent with the dimensional analysis (see Eq.\eqref{eq:ci-naive-estimate}), is independence of the suppression effect of the dark matter mixing parameter on the scale $||M_R||$ (or $\Lambda(M_R)$ in the numerical analysis). 
Numerical scan performed for the scales 10, 100, $10^3$ and $10^4$ TeV, demonstrated no dependence. 
Since the $M_R$-scale is associated with the emergence of New Physics at high energies, this makes possible to consider a dark matter scenario based on the inverse seesaw mechanism with the proposed diagonalization in combination with a low-energy extensions of the Standard Model in the TeV-energy range, accessible for collider experiments.

The models of the considered type do not require the use of
an enormous Majorana mass term of the order of $||M_R||\sim \mathcal{O}(10^{15} \text{ GeV})$ (as in the canonical type I seesaw). Furthermore, ISS models also hold an advantage over the $\nu$MSM \cite{Asaka:2005an} and $\nu$MSM-like models, where the Majorana mass parameter is comparable to the electroweak scale. In those models, the small mixing $\sim \frac{||m_D||}{||M_R||} \ll 1$ and sub-eV neutrino masses are achieved by making the Yukawa constants extremely small, $||Y|| \ll 10^{-6}$. In the present work the active neutrino mass operator is reproduced and an analytical form for the full mixing matrix of the $\iss{p}{q}$ model with arbitrary integer values of $\p$ and $\q$ is obtained. It is demonstrated that for the class of $\iss{p}{q}$ models there exists a hierarchy of mixing components $\theta_1 = \varepsilon(\mu, M_R) \cdot \theta_2 \ll \theta_2$, where $\varepsilon(\mu, M_R) \propto \frac{||\mu||}{||M_R||} \ll 10^{-6}$. At the same time, the mass operator achieves a small magnitude due to the natural smallness of $\theta_1$, namely $m_\nu = \theta_1 m_D^T$. 

In the general case selective suppression of the mixing parameter $U^2_{\rm DM}$ can be possible using a specific hierarchy of the Yukawa matrix $Y_R$ elements for the lagrangian terms with $\nu_R$ and $S_R$. From the perspective of the full Majorana mass matrix, this would mean that there is a certain $i$-th row (and $i$-th column, since the matrix is symmetric) with a maximum element of $\sim \epsilon \cdot ||M_R||\ll ||M_R||$
\begin{equation}
    \label{eq:M23-with-supression}
    \mathbf{M}_{(2,3)}\Big|_{\text{DM scenario}} =
    \left(
    \begin{array}{c|c}
        \mathbf{M}_{(2,2)} & \begin{array}{c} \mathbb{O}_{3\times1} \\ m_{R\,2\times1} \\ \mathbb{O}_{2\times1} \end{array} 
        \\
        \hline
        \begin{array}{ccc} \mathbb{O}_{1\times3} & m^T_{R\,1\times2} & \mathbb{O}_{1\times2} \end{array} & \mu_3
    \end{array}
    \right),
\end{equation}
where the Dirac mass terms are $(m_R)_{1,2} \sim \epsilon ||M_R||$. For $||M_R|| \simeq 10 \text{ TeV}$, these elements would be of the order of $O(1\text{ GeV})$. 

\section*{Funding}
The study was conducted under the state assignment of Lomonosov Moscow State University. The work of D.~K. was supported by the Theoretical Physics and Mathematics Advancement Foundation “BASIS” Grant No. 23-2-2-19-1.

\appendix

\section{Some elements of linear algebra}

\subsection{Schur complement} Let there be a block matrix of the form
    \begin{equation}
    \label{def:block-matrix}
        M = 
        \begin{pmatrix}
            A & B \\
            C & D
        \end{pmatrix}.
    \end{equation}
    Then, in the case where $D$ is invertible (i.e. $D^{-1}$ exists), the \textit{Schur complement} of $M$ with respect to $D$ is the matrix $(M|D)$, defined as
    \begin{equation}
     \label{def:shur-compl}
        (M|D) \equiv A - B D^{-1} C.
    \end{equation}
    In this case, matrix $A$ does not necessarily have to be invertible.
    However, if matrix $A$ is invertible while matrix $D$ is not, one can define the Schur complement of $M$ with respect to $A$, namely
    \begin{equation}
        (M|A) \equiv D - C A^{-1} B.
    \end{equation}
    See Refs.~\cite{shur_compl_1, shur_compl_2} for details.
\subsection{Frobenius formula for block matrix inversion} 
Let there be a block matrix Eq.~\eqref{def:block-matrix}
and let $D$ be an invertible matrix. Then the inverse matrix $M^{-1}$ has the form 
\begin{equation}
    \label{f:froben-inverse}
        M^{-1} 
        = 
        \begin{pmatrix}
            (M|D)^{-1} &  - (M|D)^{-1} B D^{-1} \\
            -D^{-1}C (M|D)^{-1} & D^{-1} + D^{-1}C (M|D)^{-1} B D^{-1} \\
        \end{pmatrix}.
\end{equation}
In the case where matrix $A$ is invertible, while matrix $D$ is not necessarily invertible, the formula can be rewritten as
\begin{equation}
\label{f:froben-inverse-alt}
    M^{-1} 
    = 
    \begin{pmatrix}
    A^{-1}+A^{-1} B (M|A)^{-1} C A^{-1} & - A^{-1} B (M|A)^{-1} \\
    - (M|A)^{-1} C A^{-1} & (M|A)^{-1}
    \end{pmatrix}.
\end{equation}
See Ref.~\cite{shur_compl_2} for details.


\begin{thebibliography}{10}

\bibitem{mohapatra_0}
R.~N. Mohapatra,
\newblock Mechanism for understanding small neutrino mass in superstring theories,
\newblock {\em Phys. Rev. Lett.} \textbf{56}, 561, 1986.

\bibitem{Mohapatra:2006gs}
R.~N. Mohapatra and A.~Y. Smirnov,
\newblock {Neutrino Mass and New Physics},
\newblock {\em Ann. Rev. Nucl. Part. Sci.} \textbf{56}, 569, 2006
[arXiv:hep-ph/0603118].

\bibitem{mohapatra_2}
R.~N. Mohapatra and J.~W.~F. Valle,
\newblock {Neutrino Mass and Baryon Number Nonconservation in Superstring Models},
\newblock {\em Phys. Rev. D} \textbf{34}, 1642, 1986.

\bibitem{pdg2024}
S.~Navas et~al., 
\newblock {Review of particle physics},
\newblock {\em Phys. Rev. D} \textbf{110}, no 3, 030001, 2024.

\bibitem{gut_iss_1}
M.~Malinsk\'y, J.~C. Rom\~ao, and J.~W.~F. Valle,
\newblock Novel supersymmetric $SO(10)$ seesaw mechanism,
\newblock {\em Phys. Rev. Lett.} \textbf{95}, 161801, 2005
[arXiv:hep-ph/0506296].

\bibitem{gut_iss_2}
R.~L. Awasthi and M.~K. Parida,
\newblock Inverse seesaw mechanism in nonsupersymmetric $SO(10)$, proton lifetime, nonunitarity effects, and a low-mass ${Z}^{\ensuremath{'}}$ boson.
\newblock {\em Phys. Rev. D} \textbf{86}, 093004, 2012
[arXiv:1112.1826 [hep-ph]].

\bibitem{Abada:2012cq}
A. Abada, D. Das, A. Vicente, and C. Weiland,
\newblock {Enhancing lepton flavour violation in the supersymmetric inverse seesaw beyond the dipole contribution},
\newblock {\em JHEP} \textbf{09}, 015, 2012
[arXiv:1206.6497 [hep-ph]].

\bibitem{Johnson:1986ux}
R. Johnson, S. Ranfone, and J. Schechter,
\newblock {Fritzsch-Stech model in SO(10)},
\newblock {\em Phys. Rev. D} \textbf{35}, 282, 1987, 
https://doi.org/10.1103/PhysRevD.35.282.

\bibitem{Glashow:1990dg} 
S.~L. Glashow,
\newblock {A novel neutrino mass hierarchy},
\newblock {\em Phys. Lett. B} \textbf{256}, 255, 1991,
https://doi.org/10.1016/0370-2693(91)90683-H.

\bibitem{Gelmini:1994az}
G. Gelmini and E. Roulet, 
\newblock {Neutrino masses},
\newblock {\em Rept. Prog. Phys.} \textbf{58}, 1207, 1995,
https://doi.org/10.1088/0034-4885/58/10/002
[arXiv:hep-ph/9412278].

\bibitem{Abada:2014kba}
A. Abada, M.~E. Krauss, W. Porod, F. Staub,  A. Vicente, and C. Weiland,
\newblock {Lepton flavor violation in low-scale seesaw models: SUSY and non-SUSY contributions},
\newblock {\em JHEP} \textbf{11}, 048, 2014,
https://doi.org/10.1007/JHEP11(2014)048 
[arXiv:1408.0138 [hep-ph]].

\bibitem{Abada_2017}
A. Abada, G. Arcadi, V. Domcke, and M. Lucente,
\newblock Neutrino masses, leptogenesis and dark matter from small lepton number violation?
\newblock {\em JCAP} \textbf{2017}, no 12, 024, 2017
[arXiv:1709.00415 [hep-ph]].

\bibitem{Lucente_2014}
A. Abada and M. Lucente,
\newblock {Looking for the minimal inverse seesaw realisation},
\newblock {\em Nuclear Physics B} \textbf{885}, 651, 2014,
https://doi.org/10.1016/j.nuclphysb.2014.06.003
[arXiv:1401.1507 [hep-ph]].

\bibitem{Abada_wdm}
A. Abada, G. Arcadi, and M. Lucente,
\newblock Dark matter in the minimal inverse seesaw mechanism,
\newblock {\em Journal of Cosmology and Astroparticle Physics} \textbf{2014}, no 10, 001, 2014
[arXiv:1406.6556 [hep-ph]].

\bibitem{Fukugita:1986hr}
M.~Fukugita and T.~Yanagida,
\newblock {Baryogenesis Without Grand Unification},
\newblock {\em Phys. Lett. B} \textbf{174}, 45, 1986.

\bibitem{Kuzmin:1985mm}
V.~A. Kuzmin, V.~A. Rubakov, and M.~E. Shaposhnikov,
\newblock {On the Anomalous Electroweak Baryon Number Nonconservation in the Early Universe},
\newblock {\em Phys. Lett. B} \textbf{155}, 36, 1985.

\bibitem{Ibarra:2010xw}
A.~Ibarra, E.~Molinaro, and S.~T. Petcov,
\newblock {TeV Scale See-Saw Mechanisms of Neutrino Mass Generation, the Majorana Nature of the Heavy Singlet Neutrinos and $(\beta\beta)_{0\nu}$-Decay},
\newblock {\em JHEP} \textbf{09}, 108, 2010
[arXiv:1007.2378 [hep-ph]].

\bibitem{own_jetp}
M.~N. Dubinin and D.~M. Kazarkin,
\newblock {Improved Cosmological Bounds for Mixing Scenarios of Three Sterile Neutrino Generations},
\newblock {\em J. Exp. Theor. Phys.} \textbf{137}, no 6, 814, 2023.

\bibitem{own_sym}
M. Dubinin and E. Fedotova,
\newblock {Non-Minimal Approximation for the Type-I Seesaw Mechanism},
\newblock {\em Symmetry} \textbf{15}, no 3, 679, 2023
[arXiv:2303.06680 [hep-ph]].

\bibitem{Horn_Johnson_book}
R.~A. Horn and C.~R. Johnson,
\newblock {\em Matrix Analysis},
\newblock Cambridge University Press, 1985.

\bibitem{shur_compl_1}
Z.~Zhang,
\newblock {\em The Schur Complement and Its Applications},
\newblock Springer New York, NY, 2005.

\bibitem{shur_compl_2}
J. Gallier,
\newblock {\em Schur Complements and Applications}, 
\newblock Springer New York, NY, 2011.

\bibitem{Duka:1999uc}
P.~Duka, J.~Gluza, and M.~Zralek,
\newblock {Quantization and renormalization of the manifest left-right symmetric model of electroweak interactions},
\newblock {\em Annals Phys.} \textbf{280}, 336, 2000 
[arXiv:hep-ph/9910279].

\bibitem{Grimus:1989pu}
W.~Grimus and H.~Neufeld,
\newblock {Radiative Neutrino Masses in an $SU(2) \times U(1)$ Model},
\newblock {\em Nucl. Phys. B} \textbf{325}, 18, 1989.

\bibitem{planck-2018}
N. Aghanim et al.,
\newblock {Planck 2018 results. VI. Cosmological parameters},
\newblock {\em Astron. Astrophys.} \textbf{641}, A6, 2020
[Erratum: Astron.Astrophys. 652, C4 (2021)],
https://doi.org/10.1051/0004-6361/201833910
[arXiv:1807.06209 [astro-ph.CO]].

\bibitem{pseudo-dirac}
G. Dutta and A.~S. Joshipura,
\newblock Pseudo dirac neutrinos in the seesaw model,
\newblock {\em Phys. Rev. D} \textbf{51}, 3838, 1995
[arXiv:hep-ph/9405291].

\bibitem{Jarlskog:1998uf}
C. Jarlskog, M. Matsuda, S. Skadhauge, and M. Tanimoto,
\newblock {Zee mass matrix and bimaximal neutrino mixing},
\newblock {\em Phys. Lett. B} \textbf{448}, 240, 1999
[arXiv:hep-ph/9812282].

\bibitem{svd_method}
G.~H. Golub and W.~Kahan,
\newblock {Calculating the singular values and pseudo-inverse of a matrix},
\newblock {\em Journal of the Society for Industrial and Applied Mathematics, Series B: Numerical Analysis} \textbf{2}, 205, 1965.

\bibitem{wolfram_SVD}
Wolfram-Research,
\newblock {SingularValueDecomposition}.
\newblock \url{https://reference.wolfram.com/language/ref/ \\ SingularValueDecomposition.html}, 2025.

\bibitem{vysotsky}
T.~M. Aliev and M. I. Vysotskii,
\newblock {Prospects for detecting photons produced by the decay of primordial neutrinos in the universe},
\newblock {\em Soviet Physics Uspekhi} \textbf{24}, 1008, 1981,
https://doi.org/10.1070/PU1981v024n12ABEH004759.

\bibitem{xray-Wolfenstein}
P.~B.Pal and L. Wolfenstein, 
\newblock {Radiative decays of massive neutrinos},
\newblock {\em Phys. Rev. D} \textbf{25}, 766, 1982,
https://link.aps.org/doi/10.1103/PhysRevD.25.766. 

\bibitem{xray-Abazajian}
K. Abazajian, G.~M. Fuller, and W.~H. Tucker,
\newblock {Direct detection of warm dark matter in the X-ray}",
\newblock {\em Astrophys. J.} \textbf{562}, 593, 2001,
[arXiv:astro-ph/0106002].

\bibitem{nuSTAR}
B.~M. Roach et al., 
\newblock {Long-exposure NuSTAR constraints on decaying dark matter in the Galactic halo},
\newblock {\em Phys. Rev. D} \textbf{107}, no 2, 023009, 2023
[arXiv:2207.04572 [astro-ph.HE]].

\bibitem{XMM}
J.~W. Foster et al.,
\newblock {Deep Search for Decaying Dark Matter with XMM-Newton Blank-Sky Observations},
\newblock {\em Phys. Rev. Lett.} \textbf{127}, no 5, 051101, 2021
[arXiv:2102.02207 [astro-ph.CO]].

\bibitem{erosita}
J.~T. Calvo, M. Taoso, A. Caputo, M. Negro, and M. Regis,
\newblock {Searching for dark matter X-ray lines from the Large Magellanic Cloud with eROSITA},
arXiv:2603.19109 [hep-ph].

\bibitem{athena}
A. Neronov and D. Malyshev,
\newblock {Toward a full test of the $\nu$MSM sterile neutrino dark matter model with Athena},
\newblock {\em Phys. Rev. D} \textbf{93}, no 6, 0635181, 2016
[arXiv:1509.02758 [astro-ph.HE]].

\bibitem{Asaka:2005an}
T. Asaka, S. Blanchet, and M. Shaposhnikov,
\newblock {The $\nu$MSM, dark matter and neutrino masses},
\newblock {\em Phys. Lett. B} \textbf{631}, 151, 2005
[arXiv:hep-ph/0503065].


\end{thebibliography}


\end{document}